\documentclass[twocolumn,prd,showpacs,preprintnumbers,amsmath,amssymb]{revtex4}
\usepackage{amsmath,mathrsfs,amsbsy,color,graphicx,bm,amsthm,amsfonts}
\usepackage{units}
\usepackage{bbm}
\usepackage{times}
\usepackage{dcolumn}
\usepackage{mathrsfs}
\usepackage{amsmath,amssymb,epsfig}
\usepackage{amsmath}
\newcommand{\udots}{\mathinner{\mskip1mu\raise1pt\vbox{\kern7pt\hbox{.}}
\mskip2mu\raise4pt\hbox{.}\mskip2mu\raise7pt\hbox{.}\mskip1mu}}

\begin{document}

\title{ Quantum coherence and interference in Young's experiments}
\author{Hao-Sheng Zeng$^1$\footnote{Corresponding author: hszeng@hunnu.edu.cn}, Wen-Jing Peng$^1$, Shu-Min Wu$^2$}
\affiliation{$^1$ Department of Physics, Hunan Normal University, Changsha 410081, China\\
$^2$ Department of Physics, Liaoning Normal University, Dalian 116029, China}%

\begin{abstract}
We propose the concept of pair-wise coherence to study the relation between the $l_1$ norm of coherence and the quantum interference in Young's multi-photon multi-path experiments, where the input photons may be entangled each other. We find that only the local coherence of each single photon can make quantum interference and the collective coherence between photons has no contribution to quantum interference. The visibility of interference fringe is commonly less than the $l_1$ norm of coherence of the corresponding input state, suggesting that the $l_1$ norm of coherence is only the necessary but not sufficient condition for quantum interference. We also find that the maximal fringe visibility can reach one. The optimal input states for producing the maximal visibility are presented.
\end{abstract}
\maketitle

\section{Introduction}
\label{sec1}

Quantum coherence is one of the key features of quantum world and responsible for quantum interference and the transition from quantum to classical\cite{Zurek}. It plays an important role in quantum optics \cite{L15,L16}, quantum thermodynamics \cite{L13,L14}, quantum biology \cite{L11,L12}, and quantum information science \cite{L17,L18}.

In the realm of quantum optics, the quantum coherence of optical fields can be described in terms of phase space distributions and multipoint correlation functions\cite{Glauber1963,Sudarshan1963,Mandel1965}. However, quantum coherence is not restricted only to the optical fields. With the applications of quantum coherence and multipartite entanglement of physical particles in modern quantum technologies such as quantum-enhanced metrology and communication protocols, the viewpoint that quantum coherence is a kind of resources has been emerged \cite{Baumgratz2014,Vicente2017,Marvian2016,Marvian20161,Pires2018,Streltsov2017,Castellini}. In the pioneering work \cite{Baumgratz2014}, Baumgratz, Cramer and Plenio established a rigorous framework for the quantification of coherence from the point of resource theory. By introducing the set of incoherent states and the set of incoherent operations, as well as assuming that coherence is nonincreasing under incoherent operations, the framework of resource theory for coherence is established, in which coherence is regarded as a resource relative to the incoherent operations. Any function that maps states to the nonnegative real numbers and satisfies the requirement of nonincreasing under incoherent operations constitutes a measure of coherence. Especially, the measure of $l_1$ norm of coherence has been proposed: Given some preferred basis $\left\{
{\left| i \right\rangle } \right\}_{i = 1, \ldots ,d}$ of a
$d$-dimensional quantum system, the $l_1$ norm of coherence is defined as the sum of the absolute value of all the off-diagonal elements of the system density matrix,
\begin{equation}\label{introduction}
    C_{l_{1}}(\rho)=\sum_{i\neq j}|\rho_{i,j}|.
\end{equation}
Note that the $l_1$ norm of coherence depends on the choice of the reference basis. A given quantum state may have different values of coherence under different reference basis. In practice, the reference basis may be
dictated by the physics of the problem under consideration. For example, one may focus on the energy eigenbasis when addressing coherence in the transport phenomena and thermodynamics. In this paper, we discuss the quantum description of Young's interference experiments, the path basis is favorable.

Based on the quantification of quantum coherence, many researches about the dynamics of coherence have been done, including the freezing\cite{Bromley2015,Yuxiao2016,Silva2016}, the spreading\cite{Pozzobom2017},
sudden change\cite{Meng2020} and quantum beat\cite{Zeng2020} for quantum coherence under the influences of environments.
The distribution of quantum coherence in multipartite systems also has been studied \cite{Yaoyao2015,Radhakrishnan2016,Mateng,Liao2021}.

When talking about quantum coherence, people naturally remind of another concept--quantum interference. Generally speaking, quantum coherence is the necessary condition for producing quantum interference.
The famous Young's interference experiment in quantum optics is the most convincing evidence for the quantum coherence of light fields. However, the quantum-mechanical description about Young's interference is usually based on the level of a single photon\cite{Scully}.
As shown in Fig.1, a single photon is incident to the pinholes ${\rm P_{1}}$ and ${\rm P_{2}}$ on the screen ${\rm S}_{1}$. The photon state behind the screen ${\rm S}_{1}$ is generally described by the wave function
\begin{equation}\label{int1}
  c_{1}|1_{\mathbf{k}_{1}}\rangle+c_{2}|1_{\mathbf{k}_{2}}\rangle,
\end{equation}
where
$c_{1}$ and $c_{2}$ with $|c_{1}|^{2}+|c_{2}|^{2}=1$ are the superposition coefficients, and $\mathbf{k}_{1}$ and $\mathbf{k}_{2}$ denote the wave vectors for the photon traveling along the paths ${\rm P_{1}P}$ and ${\rm P_{2}P}$ respectively.
It has been shown by Young's experiments that the superposition wave function of Eq.(\ref{int1}) can make interference (appear interference fringe) on the screen ${\rm S}_{2}$. This interference originates of course from the quantum coherence of the state Eq.(\ref{int1}) (which is $2|c_{1}c_{2}|$ measured by the $l_1$ norm of coherence).  Now we present a question: If two photons are incident simultaneously to the screen ${\rm S}_{1}$, then the photon state after passing through the pinholes is generally described by
\begin{equation}\label{int2}
  c_{1}|2_{\mathbf{k}_{1}}\rangle+c_{2}|1_{\mathbf{k}_{1}}\rangle|1_{\mathbf{k}_{2}}\rangle+c_{3}|2_{\mathbf{k}_{2}}\rangle,
\end{equation}
with $\sum^{3}_{j=1}|c_{j}|^{2}=1$. Are we sure we can observe quantum interference on the screen ${\rm S}_{2}$? Or more generally, for $n$ ($n\geq 2$) simultaneously incident photons, what changes will happen in the quantum interference?

According to the viewpoint of reference \cite{Radhakrishnan2016},
quantum coherence in multipartite systems may be classified into the local or collective. The local coherence belongs to each single subsystem, which comes from the coherent superposition between the levels of the particular subsystem,
and the collective coherence denotes those that cannot be attributed to particular subsystems. In this sense, the coherence in state Eq.(\ref{int1}) is obviously local because it belongs to a single photon, while the state of Eq.(\ref{int2}) may have both local and collective coherence generally. In this paper, we show that the local coherence and collective coherence behave differently in Young's interference experiments: The local coherence can make quantum interference but the collective coherence can not.
In addition, we also discuss the maximal visibility of the interference fringe in the multi-photon and multi-path Young's experiment and look for the optimal input state for producing the maximal visibility.

The paper is organized as follows. In Sec.\ref{secaa}, we firstly discuss the local and collective coherence for identical particle. Then in Sec.\ref{sec2}, we discuss Young's double-path interference. In Sec.\ref{sec3}-\ref{sec5}, we extend the relevant discussions to the multi-path interference. Finally, we give the conclusions in Sec.\ref{sec6}.

\section{local and collective coherence for identical particles}
\label{secaa}

For the set of distinguishable particles, the classification of local and collective coherence has been proposed \cite{Radhakrishnan2016}. For the set of identical particles, however, no method has been established.
We here propose a method for distinguishing the local and collective coherence for the system of identical bosons (identical photons), which is based on the most basic attribute of coherence--quantum coherence originates from coherent superposition of quantum states. This attribute also decides that the method is valid only for quantum pure states.

For the bosonic system of identical particles, the wave function is symmetrical with respect to the exchange of particles. Now we consider a system of $N$ photons, which is in the pure superposition state,
\begin{equation}\label{aa1}
    |\psi\rangle=c_{1}|\psi_{1}\rangle+c_{2}|\psi_{2}\rangle+\cdots+c_{n}|\psi_{n}\rangle,
\end{equation}
where the complex coefficients fulfill the normalization $\sum_{j=1}^{n}|c_{j}|^{2}=1$, and $|\psi_{j}\rangle$ with $j=1,\ldots, n$ denote the multi-mode number states for the system of $N$ photons. For example, for the two-mode $N$-photon system, each $|\psi_{j}\rangle$ has the form $|n_{1}\rangle_{\nu_{1}}|n_{2}\rangle_{\nu_{2}}$, meaning that $n_{i}$ ($i=1,2$ and $n_{1}+n_{2}=N$ ) photons inhabit in the mode $\nu_{i}$. It is worthwhile to point out that we have not assigned label to each photon. The state $|n_{1}\rangle_{\nu_{1}}|n_{2}\rangle_{\nu_{2}}$ only means there are $n_{1}$ photons in mode $\nu_{1}$ and $n_{2}$ photons in mode $\nu_{2}$, but we do not know which $n_{1}$ photons inhabit in mode $\nu_{1}$ and which $n_{2}$ photons inhabit in mode $\nu_{2}$. For $n_{1}=3$, we can equivalently express $|3\rangle_{\nu_{1}}$ as $|3\rangle_{\nu_{1}}=|2\rangle_{\nu_{1}}|1\rangle_{\nu_{1}}=|1\rangle_{\nu_{1}}|1\rangle_{\nu_{1}}|1\rangle_{\nu_{1}}$. In this manner, the wave function of Eq.(\ref{aa1}) is clearly symmetrical with respect to photon exchange.

Eq.(\ref{aa1}) contains $n$ superposition terms, which has the total coherence $\sum_{i,j=1}^{n}|c_{i}c_{j}|$ according to the measure of $l_1$ norm of coherence. We can split the total coherence as the sum of all the pair-wise coherence in the following manner. Note that Eq.(\ref{aa1}) can make up of $C_{n}^{2}=n(n-1)/2$ pair-wise superpositions. The pair-wise superposition $c_{i}|\psi_{i}\rangle+c_{j}|\psi_{j}\rangle$ has coherence $2|c_{i}c_{j}|$. Obviously, the total coherence of $|\psi\rangle$ is equal to the sum of the coherences of its all pair-wise superpositions.

Now the pair-wise coherence can be further distinguished into the local and collective. To be specific, we take the two-photon two-path wave function of Eq.(\ref{int2}) as example. It can form three pair-wise superpositions: $c_{1}|2_{\mathbf{k}_{1}}\rangle+c_{2}|1_{\mathbf{k}_{1}}\rangle|1_{\mathbf{k}_{2}}\rangle$, $c_{2}|1_{\mathbf{k}_{1}}\rangle|1_{\mathbf{k}_{2}}\rangle+c_{3}|2_{\mathbf{k}_{2}}\rangle$ and $c_{1}|2_{\mathbf{k}_{1}}\rangle+c_{3}|2_{\mathbf{k}_{2}}\rangle$. Writing the former two superpositions as $|1_{\mathbf{k}_{1}}\rangle(c_{1}|1_{\mathbf{k}_{1}}\rangle+c_{2}|1_{\mathbf{k}_{2}}\rangle)$ and $|1_{\mathbf{k}_{2}}\rangle(c_{2}|1_{\mathbf{k}_{1}}\rangle+c_{3}|1_{\mathbf{k}_{2}}\rangle)$ respectively, we find that the coherences for these two pair-wise superpositions are local. Because except for one photon with definite probability is in the mode $\mathbf{k}_{1}$  (for the first superposition) or $\mathbf{k}_{2}$ (for the second superposition), another photon
is in the superposition between modes $\mathbf{k}_{1}$ and $\mathbf{k}_{2}$. In other words, the coherences belong to one single photon. Conversely, the coherence formed by the third pair-wise superposition is obviously collective, because it is a two-photon entangled state in the case of $c_{1}, c_{2}\neq 0$.

Based on the concept of pair-wise coherence, we can find the following results. Given a pure superposition state of definite number of photons, if all its pair-wise superpositions have only local coherence, then the state also has only local coherence; If all its pair-wise superpositions have only collective coherence, then the state also has only collective coherence; If some pair-wise superpositions have local coherence and some pair-wise superpositions have collective coherence, then the state also has both kinds of coherences. For example, the $L$-path one-photon state $c_{1}|1_{\mathbf{k}_{1}}\rangle+c_{2}|1_{\mathbf{k}_{2}}\rangle+\cdots+c_{L}|1_{\mathbf{k}_{L}}\rangle$ has only local coherence, the $L$-path two-photon state $c_{1}|2_{\mathbf{k}_{1}}\rangle+c_{2}|2_{\mathbf{k}_{2}}\rangle+\cdots+c_{L}|2_{\mathbf{k}_{L}}\rangle$ has only collective coherence, and the state of Eq.(\ref{int2}) has both local and collective coherences.

Note that the above concept of pair-wise coherences is valid only for the pure superposition states of $N$ identical particles and under the measure of the $l_1$ norm of coherence. In the following sections, we will use the notion of pair-wise coherence to study Young's quantum interference.

\section{Double-path interference}
\label{sec2}
Young's double-path experiment is one of the most classical experiments that exhibit the coherence of light. A beam of monochromatic light is split by the pinholes ${\rm P_{1}}$ and ${\rm P_{2}}$ on the opaque screen ${\rm S}_{1}$ and then re-converge at the ${\rm P}$ on the screen ${\rm S}_{2}$ (see Fig.1). For the linearly polarized light, the corresponding electric field operator can be separated into the sum of its positive and negative parts,
$E(\mathbf{r}, t)=E^{(+)}(\mathbf{r}, t)+E^{(-)}(\mathbf{r}, t)$, with
\begin{equation}\label{eq2}
  \left\{
   \begin{array}{l}
   E^{(+)}(\mathbf{r}, t)=\sum_{j=1}^{2}\varepsilon_{\mathbf{k}_{j}}a_{\mathbf{k}_{j}}e^{{\rm i}(\mathbf{k}_{j}\cdot\mathbf{r}-\nu_{\mathbf{k}_{j}}t)},  \\
   E^{(-)}(\mathbf{r}, t)=\sum_{j=1}^{2}\varepsilon_{\mathbf{k}_{j}}a^{\dagger}_{\mathbf{k}_{j}}e^{-{\rm i}(\mathbf{k}_{j}\cdot\mathbf{r}-\nu_{\mathbf{k}_{j}}t)},
     \end{array}
   \right.
  \end{equation}
where $a_{\mathbf{k}_{j}}$ and $a^{\dagger}_{\mathbf{k}_{j}}$ are the annihilation and creation operators for the light fields traveling along ${\rm P}_{1}{\rm P}$ and ${\rm P}_{2}{\rm P}$ respectively, and $\varepsilon_{\mathbf{k}_{j}}$ are the dimensions of the corresponding electric fields. Actually, $\varepsilon_{\mathbf{k}_{j}}$ depends only on the magnitude of wave vector $\mathbf{k}_{j}$ and not on its direction. Thus we may set $\varepsilon_{\mathbf{k}_{j}}=\varepsilon_{0}$ for convenience. The light intensity at the point ${\rm P}$ with position vector $\mathbf{r}$ may be written as,
\begin{equation}\label{eq3}
  I(\mathbf{r},t)=\langle\psi|E^{(-)}(\mathbf{r},t)E^{(+)}(\mathbf{r},t)|\psi\rangle,
\end{equation}
where $|\psi\rangle$ is the wave function for the light after passing through the pinholes ${\rm P_{1}}$ and ${\rm P_{2}}$. The intensity $ I(\mathbf{r},t)$ will vary with the position $\mathbf{r}$ of $P$ on the screen, i.e., appears interference fringe on the screen. One can define the so-called visibility
\begin{equation}\label{eq4}
  V=\frac{I_{\rm max}-I_{\rm min}}{I_{\rm max}+I_{\rm min}}
\end{equation}
to describe the sharpness of the interference fringe, where $I_{\rm max}$ and $I_{\rm min}$ denote the maximum and minimum of the intensity of the interference fringe.
\begin{figure}
\vspace{-2cm}
\hspace{-1.0cm}
\includegraphics[width=4.0in,height=16cm]{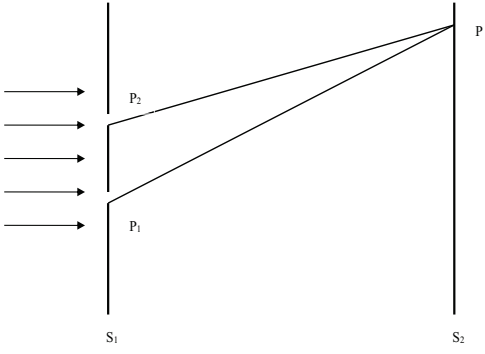}
\vspace{-9.5cm}
\caption{ Schematic diagram of Young's double-path experiment. }
\end{figure}

\subsection{One-photon interference}
\label{sec2a}
Assume that a single photon impinges on the screen ${\rm S}_{1}$, then the wave function after the pinholes ${\rm P}_{1}$ and ${\rm P}_{2}$ can be generally written as
\begin{equation}\label{eq5}
  |\psi_{2}^{1}\rangle=c_{1}|1_{\mathbf{k}_{1}}\rangle+c_{2}|1_{\mathbf{k}_{2}}\rangle,
\end{equation}
where the complex coefficients $c_{1}$ and $c_{2}$ with $|c_{1}|^{2}+|c_{2}|^{2}=1$ depend on the size and geometry of the pinholes. For this one-photon experiment, the light intensity at the screen ${\rm S}_{2}$ reads
\begin{eqnarray}\label{eq6}
 \nonumber I(\mathbf{r}) &=& \langle\psi_{2}^{1}|E^{(-)}(\mathbf{r},t)E^{(+)}(\mathbf{r},t)|\psi_{2}^{1}\rangle\\
  &=&\left|\varepsilon_{0}c_{1}e^{{\rm i}\mathbf{k}_{1}\cdot\mathbf{r}}+ \varepsilon_{0}c_{2}e^{{\rm i}\mathbf{k}_{2}\cdot\mathbf{r}}\right|^{2}\\
  \nonumber &=& |\varepsilon_{0}|^{2}+2|\varepsilon_{0}|^{2}|c_{1}c_{2}|\cos[(\mathbf{k}_{1}-\mathbf{k}_{2})\cdot\mathbf{r}+\delta_{1}-\delta_{2}],
\end{eqnarray}
where we denote $c_{j}=|c_{j}|e^{{\rm i}\delta_{j}}$ with $j=1,2$. Intensity $I(\mathbf{r})$ varies with the position ${\rm P}$ on screen ${\rm S}_{2}$,
which has the maximum $I_{\rm max}=|\varepsilon_{0}|^{2}+2|\varepsilon_{0}|^{2}|c_{1}c_{2}|$ and the minimum $I_{\rm min}=|\varepsilon_{0}|^{2}-2|\varepsilon_{0}|^{2}|c_{1}c_{2}|$ respectively.
Thus the visibility of interference fringe is
\begin{equation}\label{eq7}
  V_{2}^{1}=2|c_{1}c_{2}|.
\end{equation}
According to Eq.(\ref{introduction}), this is just the quantum coherence for the $l_1$ norm measure of the state $|\psi_{2}^{1}\rangle$. Thus for one-photon two-path interference, the fringe visibility is a good manifestation of quantum coherence of the input photon state. Obviously, when $|c_{1}|=|c_{2}|=1/\sqrt{2}$, i.e., for the equal weight superposition of the two paths of the single photon, the visibility has the maximal value $V_{2}^{1}=1$.

\subsection{Two-photon interference}
\label{sec2b}
If two photons simultaneously impinge on the screen ${\rm S}_{1}$, then the wave function after the pinholes ${\rm P}_{1}$ and ${\rm P}_{2}$ may be generally written as
\begin{equation}\label{eq8}
  |\psi_{2}^{2}\rangle=c_{1}|2_{\mathbf{k}_{1}}\rangle+c_{2}|1_{\mathbf{k}_{1}}\rangle|1_{\mathbf{k}_{2}}\rangle+c_{3}|2_{\mathbf{k}_{2}}\rangle,
\end{equation}
where $|2_{\mathbf{k}_{1}}\rangle$ and $|2_{\mathbf{k}_{2}}\rangle$ denote the states that both the two photons pass trough the pinhole ${\rm P}_{1}$ or pinhole ${\rm P}_{2}$, and $|1_{\mathbf{k}_{1}}\rangle|1_{\mathbf{k}_{2}}\rangle$ denotes the state that one photon passes through pinhole ${\rm P}_{1}$ and another photon through pinhole ${\rm P}_{2}$. The normalization condition requires $|c_{1}|^{2}+|c_{2}|^{2}+|c_{3}|^{2}=1$. The light intensity at the screen ${\rm S}_{2}$ now becomes
\begin{eqnarray}\label{eq9}
 \nonumber I(\mathbf{r}) &=& \left|\sqrt{2}\varepsilon_{0}c_{1}e^{{\rm i}\mathbf{k}_{1}\cdot\mathbf{r}}+ \varepsilon_{0}c_{2}e^{{\rm i}\mathbf{k}_{2}\cdot\mathbf{r}}\right|^{2}\\
 \nonumber &+&\left|\sqrt{2}\varepsilon_{0}c_{3}e^{{\rm i}\mathbf{k}_{2}\cdot\mathbf{r}}+ \varepsilon_{0}c_{2}e^{{\rm i}\mathbf{k}_{1}\cdot\mathbf{r}}\right|^{2}\\
  \nonumber &=& 2|\varepsilon_{0}|^{2}+2\sqrt{2}|\varepsilon_{0}|^{2}|c_{1}c_{2}|\cos[(\mathbf{k}_{1}-\mathbf{k}_{2})\cdot\mathbf{r}+\delta_{1}-\delta_{2}]\\
  &+&2\sqrt{2}|\varepsilon_{0}|^{2}|c_{2}c_{3}|\cos[(\mathbf{k}_{1}-\mathbf{k}_{2})\cdot\mathbf{r}+\delta_{2}-\delta_{3}],
\end{eqnarray}
where $c_{j}=|c_{j}|e^{{\rm i}\delta_{j}}$ with $j=1,2,3$. Under the phase match condition of $\delta_{1}-\delta_{2}=\delta_{2}-\delta_{3}({\rm mod}2\pi)$, we have
\begin{equation}\label{eq10}
  \left\{
   \begin{array}{l}
   I_{\rm max}=|\varepsilon_{0}|^{2}[2+2\sqrt{2}|c_{1}c_{2}|+2\sqrt{2}|c_{2}c_{3}|],  \\
   I_{\rm min}=|\varepsilon_{0}|^{2}[2-2\sqrt{2}|c_{1}c_{2}|-2\sqrt{2}|c_{2}c_{3}|].
     \end{array}
   \right.
  \end{equation}
The visibility of the interference fringe for this case is given by
\begin{equation}\label{eq11}
  V_{2}^{2}=\sqrt{2}[|c_{1}c_{2}|+|c_{2}c_{3}|].
\end{equation}
From this deduction, we can conclude three results:

1) Eq.(\ref{eq11}) includes only the combination $|c_{1}c_{2}|$ and $|c_{2}c_{3}|$, and not concludes the combination $|c_{1}c_{3}|$, meaning that the superpositions $c_{1}|2_{\mathbf{k}_{1}}\rangle+c_{2}|1_{\mathbf{k}_{1}}\rangle|1_{\mathbf{k}_{2}}\rangle$ and $c_{2}|1_{\mathbf{k}_{1}}\rangle|1_{\mathbf{k}_{2}}\rangle+c_{3}|2_{\mathbf{k}_{2}}\rangle$ can make interference, but the superposition $c_{1}|2_{\mathbf{k}_{1}}\rangle+c_{3}|2_{\mathbf{k}_{2}}\rangle$ can not. Note that the former two superpositions have single-photon coherence, i.e., $$c_{1}|2_{\mathbf{k}_{1}}\rangle+c_{2}|1_{\mathbf{k}_{1}}\rangle|1_{\mathbf{k}_{2}}\rangle=|1_{\mathbf{k}_{1}}\rangle(c_{1}|1_{\mathbf{k}_{1}}\rangle+c_{2}|1_{\mathbf{k}_{2}}\rangle)$$ and $$c_{2}|1_{\mathbf{k}_{1}}\rangle|1_{\mathbf{k}_{2}}\rangle+c_{3}|2_{\mathbf{k}_{2}}\rangle=|1_{\mathbf{k}_{2}}\rangle(c_{2}|1_{\mathbf{k}_{1}}\rangle+c_{3}|1_{\mathbf{k}_{2}}\rangle),$$ but the superposition $c_{1}|2_{\mathbf{k}_{1}}\rangle+c_{3}|2_{\mathbf{k}_{2}}\rangle$ has only the collective coherence. So we conclude that only the local coherence of a single photon can make interference, and the collective coherence has no contribution to quantum interference.
In practice, if the input photons are bunching, then they would tend to pass through together either pinhole ${\rm P}_{1}$ or pinhole ${\rm P}_{2}$ and the fringe visibility would reduce or disappear.

2) The fringe visibility is generally less than the $l_1$-norm of coherence.
The $l_1$-norm coherence of the input state is $C(|\psi_{2}^{2}\rangle)=2[|c_{1}c_{2}|+|c_{2}c_{3}|+|c_{1}c_{3}|]$, which obviously fulfills $V_{2}^{2}\leq C(|\psi_{2}^{2}\rangle)$. This means that
the fringe visibility in the two-photon two-path interference may not be a good manifestation of quantum coherence of the input states. Especially for $c_{2}=0$, we have $V_{2}^{2}=0$ but $C(|\psi_{2}^{2}\rangle)=2|c_{1}c_{3}|$. The quantum coherence is only the necessary but not sufficient condition for quantum interference.

3) Using the Lagrange multiplier (see Appendix A), we find that when $|c_{1}|=|c_{3}|=1/2$ and $|c_{2}|=1/\sqrt{2}$, the visibility reaches its maximal value $V_{2}^{2}=1$.

\subsection{Three-photon interference}
\label{sec2c}
For the three-photon Young's two-path experiments, the input wave function of the light fields may be generally written as
\begin{equation}\label{eq12}
  |\psi_{2}^{3}\rangle=c_{1}|3_{\mathbf{k}_{1}}\rangle+c_{2}|2_{\mathbf{k}_{1}}\rangle|1_{\mathbf{k}_{2}}\rangle
  +c_{3}|1_{\mathbf{k}_{1}}\rangle|2_{\mathbf{k}_{2}}\rangle+c_{4}|3_{\mathbf{k}_{2}}\rangle
\end{equation}
with $\sum_{j=1}^{4}|c_{j}|^{2}=1$. The light intensity on the screen ${\rm S}_{2}$ is
\begin{eqnarray}\label{eq13}
 \nonumber I(\mathbf{r}) &=& \left|\sqrt{3}\varepsilon_{0}c_{1}e^{{\rm i}\mathbf{k}_{1}\cdot\mathbf{r}}+ \varepsilon_{0}c_{2}e^{{\rm i}\mathbf{k}_{2}\cdot\mathbf{r}}\right|^{2}\\
 \nonumber &+&\left|\sqrt{2}\varepsilon_{0}c_{2}e^{{\rm i}\mathbf{k}_{1}\cdot\mathbf{r}}+ \sqrt{2}\varepsilon_{0}c_{3}e^{{\rm i}\mathbf{k}_{2}\cdot\mathbf{r}}\right|^{2}\\
 \nonumber &+& \left|\varepsilon_{0}c_{3}e^{{\rm i}\mathbf{k}_{1}\cdot\mathbf{r}}+ \sqrt{3}\varepsilon_{0}c_{4}e^{{\rm i}\mathbf{k}_{2}\cdot\mathbf{r}}\right|^{2}\\
  \nonumber &=& 3|\varepsilon_{0}|^{2}+2\sqrt{3}|\varepsilon_{0}|^{2}|c_{1}c_{2}|\cos[(\mathbf{k}_{1}-\mathbf{k}_{2})\cdot\mathbf{r}+\delta_{1}-\delta_{2}]\\
  \nonumber &+&4|\varepsilon_{0}|^{2}|c_{2}c_{3}|\cos[(\mathbf{k}_{1}-\mathbf{k}_{2})\cdot\mathbf{r}+\delta_{2}-\delta_{3}]\\
  &+& 2\sqrt{3}|\varepsilon_{0}|^{2}|c_{3}c_{4}|\cos[(\mathbf{k}_{1}-\mathbf{k}_{2})\cdot\mathbf{r}+\delta_{3}-\delta_{4}],
\end{eqnarray}
where $c_{j}=|c_{j}|e^{{\rm i}\delta_{j}}$ with $j=1,2,3,4$. Under the phase match condition of $\delta_{1}-\delta_{2}=\delta_{2}-\delta_{3}=\delta_{3}-\delta_{4}({\rm mod}2\pi)$, we have
\begin{equation}\label{eq14}
  \left\{
   \begin{array}{l}
   I_{\rm max}=|\varepsilon_{0}|^{2}[3+2\sqrt{3}|c_{1}c_{2}|+4|c_{2}c_{3}|+2\sqrt{3}|c_{3}c_{4}],  \\
   I_{\rm min}=|\varepsilon_{0}|^{2}[3-2\sqrt{3}|c_{1}c_{2}|-4|c_{2}c_{3}|-2\sqrt{3}|c_{3}c_{4}],
     \end{array}
   \right.
  \end{equation}
and the visibility is
\begin{equation}\label{eq15}
  V_{2}^{3}=\frac{2}{3}[\sqrt{3}|c_{1}c_{2}|+2|c_{2}c_{3}|+\sqrt{3}|c_{3}c_{4}|].
\end{equation}
From this deduction, we also find the similar results:

1) The superposition states
$c_{1}|3_{\mathbf{k}_{1}}\rangle+c_{4}|3_{\mathbf{k}_{2}}\rangle$, as well as
\begin{equation}\label{new1}
     \begin{array}{l}
   c_{1}|3_{\mathbf{k}_{1}}\rangle+c_{3}|1_{\mathbf{k}_{1}}\rangle|2_{\mathbf{k}_{2}}\rangle=|1_{\mathbf{k}_{1}}\rangle(c_{1}|2_{\mathbf{k}_{1}}\rangle+c_{3}|2_{\mathbf{k}_{2}}\rangle),  \\
   c_{2}|2_{\mathbf{k}_{1}}\rangle|1_{\mathbf{k}_{2}}\rangle+c_{4}|3_{\mathbf{k}_{2}}\rangle=|1_{\mathbf{k}_{2}}\rangle(c_{2}|2_{\mathbf{k}_{1}}\rangle+c_{4}|2_{\mathbf{k}_{2}}\rangle),
     \end{array}
    \end{equation}
have only collective coherence, and thus have not contribution to interference. The following superpositions,
\begin{equation}\label{new2}
     \begin{array}{l}
   c_{1}|3_{\mathbf{k}_{1}}\rangle+c_{2}|2_{\mathbf{k}_{1}}\rangle|1_{\mathbf{k}_{2}}\rangle=|2_{\mathbf{k}_{1}}\rangle(c_{1}|1_{\mathbf{k}_{1}}\rangle+c_{2}|1_{\mathbf{k}_{2}}\rangle),  \\
   c_{2}|2_{\mathbf{k}_{1}}\rangle|1_{\mathbf{k}_{2}}\rangle+c_{3}|1_{\mathbf{k}_{1}}\rangle|2_{\mathbf{k}_{2}}\rangle=|1_{\mathbf{k}_{1}}\rangle|1_{\mathbf{k}_{2}}\rangle(c_{2}|1_{\mathbf{k}_{1}}\rangle+c_{3}|1_{\mathbf{k}_{2}}\rangle),\\
   c_{3}|1_{\mathbf{k}_{1}}\rangle|2_{\mathbf{k}_{2}}\rangle+c_{4}|3_{\mathbf{k}_{2}}\rangle=|2_{\mathbf{k}_{2}}\rangle(c_{3}|1_{\mathbf{k}_{1}}\rangle+c_{4}|1_{\mathbf{k}_{2}}\rangle),
     \end{array}
    \end{equation}
have local coherence of a single photon, and thus have contributions to $V_{2}^{3}$.

2) The fringe visibility $V_{2}^{3}$ is generally less than the $l_1$-norm coherence of the state $|\psi_{2}^{3}\rangle$,
because $C(|\psi_{2}^{3}\rangle)=2[|c_{1}c_{2}|+|c_{1}c_{3}|+|c_{1}c_{4}|+|c_{2}c_{3}|+|c_{2}c_{4}|+|c_{3}c_{4}|]\geq V_{2}^{3}$.

3) If we regard $V_{2}^{3}$ as a function of the coefficients $c_{j}$ with $j=1,\ldots, 4$, from the symmetry of Eqs.(\ref{eq12}) and (\ref{eq15}),
we can affirm that the maximum of $V_{2}^{3}$ appears at the conditions $|c_{1}|=|c_{4}|$ and $|c_{2}|=|c_{3}|$. Then we can further obtain by use of the Lagrange multiplier that when
$|c_{1}|=|c_{4}|=1/\sqrt{8}$ and $|c_{2}|=|c_{3}|=\sqrt{3/8}$, the visibility has the maximal value $V_{2}^{3}=1$.

\subsection{Four-photon interference}
\label{sec2d}
For the four-photon Young's two-path interference, we write the input wave function of the light fields as
\begin{eqnarray}\label{eq16}
  \nonumber |\psi_{2}^{4}\rangle&=&c_{1}|4_{\mathbf{k}_{1}}\rangle+c_{2}|3_{\mathbf{k}_{1}}\rangle|1_{\mathbf{k}_{2}}\rangle
  +c_{3}|2_{\mathbf{k}_{1}}\rangle|2_{\mathbf{k}_{2}}\rangle\\
  &+&c_{4}|1_{\mathbf{k}_{1}}\rangle|3_{\mathbf{k}_{2}}\rangle+c_{5}|4_{\mathbf{k}_{2}}\rangle
\end{eqnarray}
with $\sum_{j=1}^{5}|c_{j}|^{2}=1$. The light intensity at the screen ${\rm S}_{2}$ is
\begin{eqnarray}\label{eq17}
 \nonumber I(\mathbf{r})
  \nonumber &=& 4|\varepsilon_{0}|^{2}+4|\varepsilon_{0}|^{2}|c_{1}c_{2}|\cos[(\mathbf{k}_{1}-\mathbf{k}_{2})\cdot\mathbf{r}+\delta_{1}-\delta_{2}]\\
  \nonumber &+&2\sqrt{6}|\varepsilon_{0}|^{2}|c_{2}c_{3}|\cos[(\mathbf{k}_{1}-\mathbf{k}_{2})\cdot\mathbf{r}+\delta_{2}-\delta_{3}]\\
  \nonumber &+& 2\sqrt{6}|\varepsilon_{0}|^{2}|c_{3}c_{4}|\cos[(\mathbf{k}_{1}-\mathbf{k}_{2})\cdot\mathbf{r}+\delta_{3}-\delta_{4}]\\
  &+& 4|\varepsilon_{0}|^{2}|c_{4}c_{5}|\cos[(\mathbf{k}_{1}-\mathbf{k}_{2})\cdot\mathbf{r}+\delta_{4}-\delta_{5}],
\end{eqnarray}
where $c_{j}=|c_{j}|e^{{\rm i}\delta_{j}}$ with $j=1,2,\ldots, 5$. Under the phase match condition of $\delta_{1}-\delta_{2}=\delta_{2}-\delta_{3}=\delta_{3}-\delta_{4}=\delta_{4}-\delta_{5}({\rm mod}2\pi)$, one can obtain the visibility of the interference fringe given by
\begin{equation}\label{eq18}
  V_{2}^{4}=\frac{1}{2}[2|c_{1}c_{2}|+\sqrt{6}|c_{2}c_{3}|+\sqrt{6}|c_{3}c_{4}|+2|c_{4}c_{5}|].
\end{equation}
Obviously, $ V_{2}^{4}$ is in general less than the $l_1$-norm coherence of the state $|\psi_{2}^{4}\rangle$ given by  $C(|\psi_{2}^{4}\rangle)=2[|c_{1}c_{2}|+|c_{1}c_{3}|+|c_{1}c_{4}|+|c_{1}c_{5}|+|c_{2}c_{3}|+|c_{2}c_{4}|+|c_{2}c_{5}|+|c_{3}c_{4}|+|c_{3}c_{5}|+|c_{4}c_{5}|]$.
The superpositions that cannot make interference are
\begin{equation}\label{new3}
     \begin{array}{l}
   c_{1}|4_{\mathbf{k}_{1}}\rangle+c_{3}|2_{\mathbf{k}_{1}}\rangle|2_{\mathbf{k}_{2}}\rangle=|2_{\mathbf{k}_{1}}\rangle(c_{1}|2_{\mathbf{k}_{1}}\rangle+c_{3}|2_{\mathbf{k}_{2}}\rangle),  \\
   c_{1}|4_{\mathbf{k}_{1}}\rangle+c_{4}|1_{\mathbf{k}_{1}}\rangle|3_{\mathbf{k}_{2}}\rangle=|1_{\mathbf{k}_{1}}\rangle(c_{1}|3_{\mathbf{k}_{1}}\rangle+c_{4}|3_{\mathbf{k}_{2}}\rangle),\\
   c_{2}|3_{\mathbf{k}_{1}}\rangle|1_{\mathbf{k}_{2}}\rangle+c_{4}|1_{\mathbf{k}_{1}}\rangle|3_{\mathbf{k}_{2}}\rangle=|1_{\mathbf{k}_{1}}\rangle|1_{\mathbf{k}_{2}}\rangle(c_{2}|2_{\mathbf{k}_{1}}\rangle+c_{4}|2_{\mathbf{k}_{2}}\rangle),\\
   c_{2}|3_{\mathbf{k}_{1}}\rangle|1_{\mathbf{k}_{2}}\rangle+c_{5}|4_{\mathbf{k}_{2}}\rangle=|1_{\mathbf{k}_{2}}\rangle(c_{2}|3_{\mathbf{k}_{1}}\rangle+c_{5}|3_{\mathbf{k}_{2}}\rangle),\\
   c_{3}|2_{\mathbf{k}_{1}}\rangle|2_{\mathbf{k}_{2}}\rangle+c_{5}|4_{\mathbf{k}_{2}}\rangle=|2_{\mathbf{k}_{2}}\rangle(c_{3}|2_{\mathbf{k}_{1}}\rangle+c_{5}|2_{\mathbf{k}_{2}}\rangle),
     \end{array}
    \end{equation}
as well as $c_{1}|4_{\mathbf{k}_{1}}\rangle+c_{5}|4_{\mathbf{k}_{2}}\rangle$, which obviously conclude only the collective coherence. Conversely, the superpositions,
\begin{equation}\label{new4}
     \begin{array}{l}
   c_{1}|4_{\mathbf{k}_{1}}\rangle+c_{2}|3_{\mathbf{k}_{1}}\rangle|1_{\mathbf{k}_{2}}\rangle=|3_{\mathbf{k}_{1}}\rangle(c_{1}|1_{\mathbf{k}_{1}}\rangle+c_{2}|1_{\mathbf{k}_{2}}\rangle),  \\
   c_{2}|3_{\mathbf{k}_{1}}\rangle|1_{\mathbf{k}_{2}}\rangle+c_{3}|2_{\mathbf{k}_{1}}\rangle|2_{\mathbf{k}_{2}}\rangle=|2_{\mathbf{k}_{1}}\rangle|1_{\mathbf{k}_{2}}\rangle(c_{2}|1_{\mathbf{k}_{1}}\rangle+c_{3}|1_{\mathbf{k}_{2}}\rangle),\\
   c_{3}|2_{\mathbf{k}_{1}}\rangle|2_{\mathbf{k}_{2}}\rangle+c_{4}|1_{\mathbf{k}_{1}}\rangle|3_{\mathbf{k}_{2}}\rangle=|1_{\mathbf{k}_{1}}\rangle|2_{\mathbf{k}_{2}}\rangle(c_{3}|1_{\mathbf{k}_{1}}\rangle+c_{4}|1_{\mathbf{k}_{2}}\rangle),\\
   c_{4}|1_{\mathbf{k}_{1}}\rangle|3_{\mathbf{k}_{2}}\rangle+c_{5}|4_{\mathbf{k}_{2}}\rangle=|3_{\mathbf{k}_{2}}\rangle(c_{4}|1_{\mathbf{k}_{1}}\rangle+c_{5}|1_{\mathbf{k}_{2}}\rangle),
     \end{array}
    \end{equation}
have one-photon local coherence, and thus have contributions to $V_{2}^{4}$. By use of the symmetry of Eqs.(\ref{eq16}) and (\ref{eq18}) as well as the Lagrange multiplier, we find that when $|c_{1}|=|c_{5}|=1/4$, $|c_{2}|=|c_{4}|=1/2$ and $|c_{3}|=\sqrt{6}/4$,
the visibility reaches its maximal value $V_{2}^{4}=1$.

\subsection{n-photon interference}
\label{sec2e}
Finally, we discuss the n-photon Young's two-path interference, for which the input wave function of the photons is written as
\begin{eqnarray}\label{eq19}
  \nonumber |\psi_{2}^{n}\rangle&=&c_{1}|n_{\mathbf{k}_{1}}\rangle+c_{2}|(n-1)_{\mathbf{k}_{1}}\rangle|1_{\mathbf{k}_{2}}\rangle
  +c_{3}|(n-2)_{\mathbf{k}_{1}}\rangle|2_{\mathbf{k}_{2}}\rangle\\
  &+&\cdots +c_{n}|1_{\mathbf{k}_{1}}\rangle|(n-1)_{\mathbf{k}_{2}}\rangle+c_{n+1}|n_{\mathbf{k}_{2}}\rangle
\end{eqnarray}
with $\sum_{j=1}^{n+1}|c_{j}|^{2}=1$. The light intensity at the screen ${\rm S}_{2}$ is given by
\begin{eqnarray}\label{eq20}
 \nonumber I(\mathbf{r})
  \nonumber = n|\varepsilon_{0}|^{2}&+&2|\varepsilon_{0}|^{2}\sum_{j=0}^{n-1}\sqrt{(n-j)(j+1)}|c_{j+1}c_{j+2}|\\
  &\times&\cos[(\mathbf{k}_{1}-\mathbf{k}_{2})\cdot\mathbf{r}+\delta_{j+1}-\delta_{j+2}],
\end{eqnarray}
where $c_{j}=|c_{j}|e^{{\rm i}\delta_{j}}$ with $j=1,2,\ldots, (n+1)$. Under the phase match condition of $\delta_{j+1}-\delta_{j+2}=\delta_{1}-\delta_{2}({\rm mod}2\pi)$ with $j=1,2, \ldots, (n-1)$, one obtains the visibility of the interference fringe,
\begin{equation}\label{eq21}
  V_{2}^{n}=\frac{2}{n}\sum_{j=0}^{n-1}\sqrt{(n-j)(j+1)}|c_{j+1}c_{j+2}|.
\end{equation}
Obviously, $ V_{2}^{n}\leq 2\sum_{j=0}^{n-1}|c_{j+1}c_{j+2}|\leq C(|\psi_{2}^{n}\rangle)$, i.e., $ V_{2}^{n}$ is in general less than the $l_1$-norm coherence of the state $|\psi_{2}^{n}\rangle$. Every superposition between the neighboring terms in Eq.(\ref{eq19}) includes the local coherence of a single photon, and thus can make contribution to  $ V_{2}^{n}$.
Any superposition between non-adjacent terms in Eq.(\ref{eq19}) can not make contribution to interference because, except for a common factor, it can be written as the form of $\alpha|j_{\mathbf{k}_{1}}\rangle+\beta|j_{\mathbf{k}_{2}}\rangle$ with $j=2,3,\ldots, n$.
As for the maximal visibility and the corresponding optimal input state, we cannot give the universal expressions. But from the preceding deductions, we have reason to believe that the maximum of $ V_{2}^{n}$ for any finite $n$ can reach one.

\section{Three-path interference}
\label{sec3}
In this and following sections, we generalize Young's interference from the double paths to multiple paths. Different from the above section that works in the Heisenberg's picture, we now switch to the Schrodinger picture.
Assume that a beam of monochromatic light is incident on an opaque screen with $L$ pinholes and then travel along $L$ different paths (see Fig.2). The light traveling along path $j$ experiences phase shift $\alpha_{j}$.
The whole phase-shift operation can be described mathematically by $P=\exp[{\rm i}\sum_{j=1}^{L}\alpha_{j}a^{\dagger}_{\mathbf{k}_{j}}a_{\mathbf{k}_{j}}]$,
where $\mathbf{k}_{j}$ denotes the wave vector of the light traveling along path $j$, and $a_{\mathbf{k}_{j}}$ is the corresponding annihilation operator.
Different from the actual Young's multi-path experiment, here each phase shift $\alpha_{j}$ can be adjusted at will independently.  Denoting the input state of the light by $|\psi\rangle_{\rm in}$, then the output state may be written as $|\psi\rangle_{\rm out}=P|\psi\rangle_{\rm in}$.
The field operators in Eq.(\ref{eq2}) are replaced now by
\begin{equation}\label{eq22}
  \left\{
   \begin{array}{l}
   E^{(+)}=\varepsilon_{0}\sum_{j=1}^{L}a_{\mathbf{k}_{j}},  \\
   E^{(-)}=\varepsilon_{0}\sum_{j=1}^{L}a^{\dagger}_{\mathbf{k}_{j}},
     \end{array}
   \right.
  \end{equation}
where $\varepsilon_{0}$ as before is the dimension of the electric field which is the same for all the paths.
The intensity of the interference fringe is thus given by
\begin{equation}\label{eq23}
  I=: _{\rm out}\langle\psi|E^{(-)}E^{(+)}|\psi\rangle_{\rm out},
\end{equation}
and the visibility of the interference fringe is calculated still through Eq.(\ref{eq4}).
\begin{figure}
\vspace{-1.5cm}
\hspace{-1.5cm}
\includegraphics[width=4.0in,height=16cm]{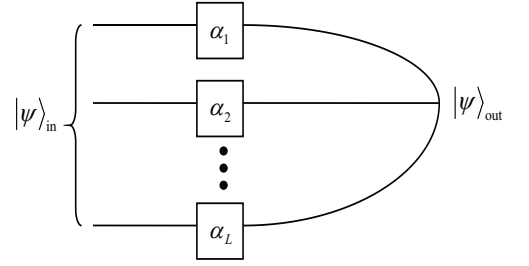}
\vspace{-11cm}
\caption{ Schematic diagram of multi-path interference. }
\end{figure}

\subsection{Three-path 1-photon interference}
\label{sec3a}
We firstly discuss the case of only a single photon, so that the input state may be written as
\begin{eqnarray}\label{eq24}
  |\psi_{3}^{1}\rangle_{\rm in}=c_{1}|1_{\mathbf{k}_{1}}\rangle+c_{2}|1_{\mathbf{k}_{2}}\rangle+c_{3}|1_{\mathbf{k}_{3}}\rangle
\end{eqnarray}
with the normalization $\sum_{j=1}^{3}|c_{j}|^{2}=1$. The output state after phase-shift operations reads
\begin{eqnarray}\label{eq25}
  |\psi_{3}^{1}\rangle_{\rm out}=c_{1}e^{{\rm i}\alpha_{1}}|1_{\mathbf{k}_{1}}\rangle+c_{2}e^{{\rm i}\alpha_{2}}|1_{\mathbf{k}_{2}}\rangle+c_{3}e^{{\rm i}\alpha_{3}}|1_{\mathbf{k}_{3}}\rangle
\end{eqnarray}
Thus the intensity of the interference fringe is
\begin{eqnarray}\label{eq26}
\nonumber I&=&|\varepsilon_{0}|^{2}\left\{1+2|c_{1}c_{2}|\cos\beta_{12}+2|c_{1}c_{3}|\cos\beta_{13}\right.\\
&+& \left. 2|c_{2}c_{3}|\cos(\beta_{13}-\beta_{12})\right\},
\end{eqnarray}
where we denote $c_{j}=|c_{j}|e^{{\rm i}\delta_{j}}$ with $j=1,2,3$, and $\beta_{ij}=(\alpha_{i}-\alpha_{j})+(\delta_{i}-\delta_{j})$. For the given input state $|\psi_{3}^{1}\rangle_{\rm in}$, we can always adjust the phase shifts
$\alpha_{j}$ to make $\beta_{12}=\beta_{13}=0$. Thus the maximum of the interference intensity is given by
\begin{equation}\label{eq27a}
  I_{\rm max}=|\varepsilon_{0}|^{2}[1+2|c_{1}c_{2}|+2|c_{1}c_{3}|+2|c_{2}c_{3}|].
\end{equation}
The minimum of the interference intensity is not so easy to find, because the three cosine functions $\cos\beta_{12}$, $\cos\beta_{13}$ and $\cos(\beta_{13}-\beta_{12})$ can not simultaneously equal $-1$. We denote $(\beta_{12}^{0},\beta_{13}^{0})$ as the position for which the interference intensity reaches its minimum, i.e.,
\begin{eqnarray}\label{eq27b}
\nonumber I_{\rm min}&=&|\varepsilon_{0}|^{2}\left\{1+2|c_{1}c_{2}|\cos\beta_{12}^{0}+2|c_{1}c_{3}|\cos\beta_{13}^{0}\right.\\
&+& \left. 2|c_{2}c_{3}|\cos(\beta_{13}^{0}-\beta_{12}^{0})\right\}.
\end{eqnarray}
Regarding $I_{\rm min}$ as a function of $\beta_{12}^{0}$ and $\beta_{13}^{0}$, according to the extremum of binary function (see Appendix B), we get
\begin{eqnarray}
\nonumber  \cos\beta_{12}^{0} &=& \frac{2|c_{3}|^{2}-1}{2|c_{1}c_{2}|}, \\
 \nonumber \cos\beta_{13}^{0} &=& - \left|\frac{2|c_{2}|^{2}-1}{2c_{1}c_{3}}\right|\\
 \nonumber \cos(\beta_{13}^{0}-\beta_{12}^{0}) &=& - \left|\frac{2|c_{1}|^{2}-1}{2c_{2}c_{3}}\right|
\end{eqnarray}
The visibility of interference fringe can be written as
\begin{widetext}
\begin{equation}\label{eq28}
  V_{3}^{1}=\frac{|c_{1}c_{2}|(1-\cos\beta_{12}^{0})+|c_{1}c_{3}|(1-\cos\beta_{13}^{0})+|c_{2}c_{3}|[1-\cos(\beta_{13}^{0}-\beta_{12}^{0})]}{1+|c_{1}c_{2}|(1+\cos\beta_{12}^{0})+|c_{1}c_{3}|(1+\cos\beta_{13}^{0})+|c_{2}c_{3}|[1+\cos(\beta_{13}^{0}-\beta_{12}^{0})]}.
\end{equation}
\end{widetext}
From the above deduction, we can find that: 1) $V_{3}^{1}$ includes the combinations $|c_{1}c_{2}|$, $|c_{1}c_{3}|$ and $|c_{2}c_{3}|$, meaning that any pair-wise superposition in Eq.(\ref{eq24}) can make interference.  Note that all these pair-wise superpositions have only single-photon coherence, we thus conclude that local coherence of a single photon can make quantum interference. 2) Note that the denominator of Eq.(\ref{eq28}) is not less than one and $1-\cos\beta_{ij}^{0}\leq 2$, thus we have $V_{3}^{1}\leq 2|c_{1}c_{2}|+2|c_{1}c_{3}|+2|c_{2}c_{3}|$, i.e., the visibility $V_{3}^{1}$ is generally less than the $l_1$-norm coherence of the input state $|\psi_{3}^{1}\rangle_{\rm in}$.
3) We find that when $c_{1}=c_{2}=c_{3}=1/\sqrt{3}$ (means $\delta_{1}=\delta_{2}=\delta_{3}=0$) and
$\alpha_{1}=4\pi/3$, $\alpha_{2}=2\pi/3$, $\alpha_{3}=0$ (means $\beta_{12}^{0}=2\pi/3$, $\beta_{13}^{0}=4\pi/3$), the visibility reaches its maximum $V_{3}^{1}=1$.

\subsection{Three-path 2-photon interference}
\label{sec3b}
Secondly, we discuss the interference of two photons traveling along three paths. We write the input state as
\begin{eqnarray}\label{eq29}
  |\psi_{3}^{2}\rangle_{\rm in}&=&c_{1}|2_{\mathbf{k}_{1}}\rangle+c_{2}|2_{\mathbf{k}_{2}}\rangle+c_{3}|2_{\mathbf{k}_{3}}\rangle\\
  \nonumber &+&c_{12}|1_{\mathbf{k}_{1}}\rangle|1_{\mathbf{k}_{2}}\rangle+c_{13}|1_{\mathbf{k}_{1}}\rangle|1_{\mathbf{k}_{3}}\rangle+c_{23}|1_{\mathbf{k}_{2}}\rangle|1_{\mathbf{k}_{3}}\rangle,
\end{eqnarray}
with the normalization $\sum_{j=1}^{3}|c_{j}|^{2}+|c_{12}|^{2}+|c_{13}|^{2}+|c_{23}|^{2}=1$. The output state after phase-shift operations reads
\begin{eqnarray}\label{eq30}
 \nonumber |\psi_{3}^{2}\rangle_{\rm out}&=&c_{1}e^{{\rm i}2\alpha_{1}}|2_{\mathbf{k}_{1}}\rangle+c_{2}e^{{\rm i}2\alpha_{2}}|2_{\mathbf{k}_{2}}\rangle+c_{3}e^{{\rm i}2\alpha_{3}}|2_{\mathbf{k}_{3}}\rangle\\
 \nonumber  &+&c_{12}e^{{\rm i}(\alpha_{1}+\alpha_{2})}|1_{\mathbf{k}_{1}}\rangle|1_{\mathbf{k}_{2}}\rangle+c_{13}e^{{\rm i}(\alpha_{1}+\alpha_{3})}|1_{\mathbf{k}_{1}}\rangle|1_{\mathbf{k}_{3}}\rangle\\
   &+&c_{23}e^{{\rm i}(\alpha_{2}+\alpha_{3})}|1_{\mathbf{k}_{2}}\rangle|1_{\mathbf{k}_{3}}\rangle.
\end{eqnarray}
The intensity of interference fringe is given by
\begin{eqnarray}\label{eq31}
 \nonumber I&=&|\sqrt{2}\varepsilon_{0}c_{1}e^{{\rm i}2\alpha_{1}}+\varepsilon_{0}c_{12}e^{{\rm i}(\alpha_{1}+\alpha_{2})}+\varepsilon_{0}c_{13}e^{{\rm i}(\alpha_{1}+\alpha_{3})}|^{2}\\
 \nonumber &+&|\sqrt{2}\varepsilon_{0}c_{2}e^{{\rm i}2\alpha_{2}}+\varepsilon_{0}c_{12}e^{{\rm i}(\alpha_{1}+\alpha_{2})}+\varepsilon_{0}c_{23}e^{{\rm i}(\alpha_{2}+\alpha_{3})}|^{2}\\
\nonumber  &+&|\sqrt{2}\varepsilon_{0}c_{3}e^{{\rm i}2\alpha_{3}}+\varepsilon_{0}c_{13}e^{{\rm i}(\alpha_{1}+\alpha_{3})}+\varepsilon_{0}c_{23}e^{{\rm i}(\alpha_{2}+\alpha_{3})}|^{2},
\end{eqnarray}
which can be equivalently written as
\begin{eqnarray}\label{eq3b2}
\nonumber   I &=& 2|\varepsilon_{0}|^{2}\left\{1+\sqrt{2}(|c_{1}c_{12}|\cos\beta_{1}+|c_{1}c_{13}|\cos\beta_{2})\right.\\
\nonumber  &+&\sqrt{2}(|c_{2}c_{12}|\cos\beta_{3}+|c_{2}c_{23}|\cos\beta_{4})\\
 &+&\sqrt{2}(|c_{3}c_{13}|\cos\beta_{5}+|c_{3}c_{23}|\cos\beta_{6}) \\
\nonumber   &+& \left. |c_{12}c_{13}|\cos\beta_{7}+|c_{12}c_{23}|\cos\beta_{8}+|c_{13}c_{23}|\cos\beta_{9} \right\},
\end{eqnarray}
where $\beta_{j}$ with $j=1,\ldots,9$ are the functions of the phase shifts $\alpha_{i}$ with $i=1,2,3$ and the arguments of the complex coefficients $c$'s in Eq.(\ref{eq29}). Especially, if we take the arguments of all the complex coefficients $c$'s to be zero, i.e.,
$\arg(c_{i})=\arg(c_{ij})=0$, then we have
\begin{eqnarray}\label{eq3b3}
\nonumber  \beta_{1}&=&\beta_{3}=\beta_{9}=\alpha_{1}-\alpha_{2}\\
\nonumber  \beta_{2}&=&\beta_{5}=\beta_{8}=\alpha_{1}-\alpha_{3}\\
  \beta_{4}&=&\beta_{6}=\beta_{7}=\alpha_{2}-\alpha_{3}.
\end{eqnarray}
Obviously $\beta_{4}=\beta_{2}-\beta_{1}$, i.e., $\beta_{j}$ are not completely independent.

For the given input state $|\psi_{3}^{2}\rangle_{\rm in}$, we can adjust the phase shifts
$\alpha_{j}$ with $j=1,2,3$ to make Eq.(\ref{eq3b2}) reaching its maximum $I_{\rm max}$ or minimum $I_{\rm min}$,
\begin{eqnarray}\label{eq3b4}
\nonumber   I_{\rm max} &=& 2|\varepsilon_{0}|^{2}\left\{1+\sqrt{2}(|c_{1}c_{12}|\cos\beta_{1}^{1}+|c_{1}c_{13}|\cos\beta_{2}^{1})\right.\\
\nonumber  &+&\sqrt{2}(|c_{2}c_{12}|\cos\beta_{3}^{1}+|c_{2}c_{23}|\cos\beta_{4}^{1})\\
 &+&\sqrt{2}(|c_{3}c_{13}|\cos\beta_{5}^{1}+|c_{3}c_{23}|\cos\beta_{6}^{1}) \\
\nonumber   &+& \left. |c_{12}c_{13}|\cos\beta_{7}^{1}+|c_{12}c_{23}|\cos\beta_{8}^{1}+|c_{13}c_{23}|\cos\beta_{9}^{1} \right\},
\end{eqnarray}
\begin{eqnarray}\label{eq3b5}
\nonumber   I_{\rm min} &=& 2|\varepsilon_{0}|^{2}\left\{1+\sqrt{2}(|c_{1}c_{12}|\cos\beta_{1}^{0}+|c_{1}c_{13}|\cos\beta_{2}^{0})\right.\\
\nonumber  &+&\sqrt{2}(|c_{2}c_{12}|\cos\beta_{3}^{0}+|c_{2}c_{23}|\cos\beta_{4}^{0})\\
 &+&\sqrt{2}(|c_{3}c_{13}|\cos\beta_{5}^{0}+|c_{3}c_{23}|\cos\beta_{6}^{0}) \\
\nonumber   &+& \left. |c_{12}c_{13}|\cos\beta_{7}^{0}+|c_{12}c_{23}|\cos\beta_{8}^{0}+|c_{13}c_{23}|\cos\beta_{9}^{0} \right\},
\end{eqnarray}
where $\{\beta_{j}^{1}\}$ and  $\{\beta_{j}^{0}\}$ with $j=1,\ldots,9$ denote respectively the positions that $I=I_{\rm max}$ and $I=I_{\rm min}$.  The visibility $V_{3}^{2}$ is then evaluated via Eq.(\ref{eq4}). Obviously, the expression of $V_{3}^{2}$
includes 9 types of combinations of the coefficients $c$'s: $|c_{1}c_{12}|$, $|c_{1}c_{13}|$, $|c_{2}c_{12}|$, $|c_{2}c_{23}|$, $|c_{3}c_{13}|$, $|c_{3}c_{23}|$, $|c_{12}c_{13}|$, $|c_{12}c_{23}|$ and $|c_{13}c_{23}|$, meaning that the corresponding 9 types of superpositions
(For example, we have the superposition $c_{1}|2_{\mathbf{k}_{1}}\rangle+c_{12}|1_{\mathbf{k}_{1}}\rangle|1_{\mathbf{k}_{2}}\rangle$ for the combination $|c_{1}c_{12}|$, etc.) can make interference. Note that all these superpositions have local coherence of a single photon,
for example, $c_{1}|2_{\mathbf{k}_{1}}\rangle+c_{12}|1_{\mathbf{k}_{1}}\rangle|1_{\mathbf{k}_{2}}\rangle=|1_{\mathbf{k}_{1}}\rangle(c_{1}|1_{\mathbf{k}_{1}}\rangle+c_{12}|1_{\mathbf{k}_{2}}\rangle)$,
and $c_{12}|1_{\mathbf{k}_{1}}\rangle|1_{\mathbf{k}_{2}}\rangle+c_{13}|1_{\mathbf{k}_{1}}\rangle|1_{\mathbf{k}_{3}}\rangle=|1_{\mathbf{k}_{1}}\rangle(c_{12}|1_{\mathbf{k}_{2}}\rangle+c_{13}|1_{\mathbf{k}_{3}}\rangle)$, and so on.
Besides these, there are other 6 types of pair-wise superpositions, $c_{1}|2_{\mathbf{k}_{1}}\rangle+c_{2}|2_{\mathbf{k}_{2}}\rangle$, $c_{1}|2_{\mathbf{k}_{1}}\rangle+c_{3}|2_{\mathbf{k}_{3}}\rangle$, $c_{2}|2_{\mathbf{k}_{2}}\rangle+c_{3}|2_{\mathbf{k}_{3}}\rangle$,
$c_{1}|2_{\mathbf{k}_{1}}\rangle+c_{23}|1_{\mathbf{k}_{2}}\rangle|1_{\mathbf{k}_{3}}\rangle$, $c_{2}|2_{\mathbf{k}_{2}}\rangle+c_{13}|1_{\mathbf{k}_{1}}\rangle|1_{\mathbf{k}_{3}}\rangle$
and $c_{3}|2_{\mathbf{k}_{3}}\rangle+c_{12}|1_{\mathbf{k}_{1}}\rangle|1_{\mathbf{k}_{2}}\rangle$. These superpositions have only collective coherence and thus have no contribution to $V_{3}^{2}$. [Note that the superposition pattern of the latter three superpositions is different from the former three!]

In order to demonstrate the fact that $V_{3}^{2}$ is generally less than the $l_1$-norm coherence of the input state $|\psi_{3}^{2}\rangle_{\rm in}$, we need the following fact:
For any $a>b>0$ and $c\geq0$, we have $\frac{b}{a}\leq\frac{b+c}{a+c}$. Denoting
\begin{eqnarray}\label{eq3b6}
\nonumber \tilde{I} &=& 2|\varepsilon_{0}|^{2}\left\{1+\sqrt{2}(|c_{1}c_{12}|+|c_{1}c_{13}|)\right.\\
\nonumber  &+&\sqrt{2}(|c_{2}c_{12}|+|c_{2}c_{23}|)\\
 &+&\sqrt{2}(|c_{3}c_{13}|+|c_{3}c_{23}|) \\
\nonumber   &+& \left. |c_{12}c_{13}|+|c_{12}c_{23}|+|c_{13}c_{23}| \right\},
\end{eqnarray}
we have $\tilde{I}+I_{\rm min}\geq 4|\varepsilon_{0}|^{2}$ and
\begin{eqnarray}\label{eq3b7}
\nonumber \tilde{I}-I_{\rm min} &\leq& 4|\varepsilon_{0}|^{2}\left\{\sqrt{2}(|c_{1}c_{12}|+|c_{1}c_{13}|)\right.\\
\nonumber  &+&\sqrt{2}(|c_{2}c_{12}|+|c_{2}c_{23}|)\\
 &+&\sqrt{2}(|c_{3}c_{13}|+|c_{3}c_{23}|) \\
\nonumber   &+& \left. |c_{12}c_{13}|+|c_{12}c_{23}|+|c_{13}c_{23}| \right\}.
\end{eqnarray}
Setting $c=\tilde{I}-I_{\rm max}$, then
\begin{eqnarray}\label{eq3b8}
\nonumber  V_{3}^{2}&=&\frac{I_{\rm max}-I_{\rm min}}{I_{\rm max}+I_{\rm min}}\leq\frac{I_{\rm max}-I_{\rm min}+c}{I_{\rm max}+I_{\rm min}+c}=\frac{\tilde{I}-I_{\rm min}}{\tilde{I}+I_{\rm min}}\\
\nonumber  &\leq& \sqrt{2}(|c_{1}c_{12}|+|c_{1}c_{13}|)\\
\nonumber  &+&\sqrt{2}(|c_{2}c_{12}|+|c_{2}c_{23}|)\\
\nonumber &+&\sqrt{2}(|c_{3}c_{13}|+|c_{3}c_{23}|) \\
\nonumber   &+& |c_{12}c_{13}|+|c_{12}c_{23}|+|c_{13}c_{23}|\\
&\leq& C(|\psi_{3}^{2}\rangle_{\rm in}).
\end{eqnarray}

Through complicated calculations, we find that the maximal visibility of $V_{3}^{2}$ in this case can also be reached one. By taking $c_{1}=c_{2}=c_{3}=1/3$, $c_{12}=c_{13}=c_{23}=\sqrt{2}/3$ (means that the arguments of all the superposition coefficients are zero),
and the phase shifts $\alpha_{1}=4\pi/3$, $\alpha_{2}=2\pi/3$, $\alpha_{3}=0$, we find that the minimal intensity of Eq.(\ref{eq3b5}) is $I_{\rm min}=0$, meaning that the visibility reaches the maximum $V_{3}^{2}=1$.

\subsection{Three-path 3-photon interference}
\label{sec3c}
For the 3-photon three-path interference, the input state may be written as
\begin{eqnarray}\label{eq35}
\nonumber  |\psi_{3}^{3}\rangle_{\rm in}&=&c_{1}|3_{\mathbf{k}_{1}}\rangle+c_{2}|3_{\mathbf{k}_{2}}\rangle+c_{3}|3_{\mathbf{k}_{3}}\rangle\\
  \nonumber &+&c_{12}|2_{\mathbf{k}_{1}}\rangle|1_{\mathbf{k}_{2}}\rangle+c_{13}|2_{\mathbf{k}_{1}}\rangle|1_{\mathbf{k}_{3}}\rangle\\
\nonumber  &+&c_{23}|2_{\mathbf{k}_{2}}\rangle|1_{\mathbf{k}_{3}}\rangle+c_{21}|2_{\mathbf{k}_{2}}\rangle|1_{\mathbf{k}_{1}}\rangle\\
\nonumber &+&c_{31}|2_{\mathbf{k}_{3}}\rangle|1_{\mathbf{k}_{1}}\rangle+c_{32}|2_{\mathbf{k}_{3}}\rangle|1_{\mathbf{k}_{2}}\rangle\\
&+& c_{33}|1_{\mathbf{k}_{1}}\rangle|1_{\mathbf{k}_{2}}\rangle|1_{\mathbf{k}_{3}}\rangle,
\end{eqnarray}
where the sum of the squares of the absolute values of all the complex coefficients equals one.
The interference intensity may be written as
\begin{equation}\label{eq3c1}
   I =3|\varepsilon_{0}|^{2}+A+A^{*}
\end{equation}
with
\begin{widetext}
\begin{eqnarray}\label{eq3c2}
    A &=&\sqrt{3}|\varepsilon_{0}|^{2}[c_{1}c_{12}^{*}e^{{\rm i}(\alpha_{1}-\alpha_{2})}+c_{1}c_{13}^{*}e^{{\rm i}(\alpha_{1}-\alpha_{3})}+c_{2}c_{23}^{*}e^{{\rm i}(\alpha_{2}-\alpha_{3})}
   +c_{2}c_{21}^{*}e^{{\rm i}(\alpha_{2}-\alpha_{1})}+c_{3}c_{31}^{*}e^{{\rm i}(\alpha_{3}-\alpha_{1})}+c_{3}c_{32}^{*}e^{{\rm i}(\alpha_{3}-\alpha_{2})}]  \\
  \nonumber  &+& \sqrt{2}|\varepsilon_{0}|^{2}[c_{12}c_{33}^{*}e^{{\rm i}(\alpha_{1}-\alpha_{3})}+c_{21}c_{33}^{*}e^{{\rm i}(\alpha_{2}-\alpha_{3})}+c_{13}c_{33}^{*}e^{{\rm i}(\alpha_{1}-\alpha_{2})}
    +c_{31}c_{33}^{*}e^{{\rm i}(\alpha_{3}-\alpha_{2})}+c_{23}c_{33}^{*}e^{{\rm i}(\alpha_{2}-\alpha_{1})}+c_{32}c_{33}^{*}e^{{\rm i}(\alpha_{3}-\alpha_{1})}]\\
 \nonumber   &+& |\varepsilon_{0}|^{2}[2c_{12}c_{21}^{*}e^{{\rm i}(\alpha_{1}-\alpha_{2})}+2c_{13}c_{31}^{*}e^{{\rm i}(\alpha_{1}-\alpha_{3})}+2c_{23}c_{32}^{*}e^{{\rm i}(\alpha_{2}-\alpha_{3})}
    +c_{12}c_{13}^{*}e^{{\rm i}(\alpha_{2}-\alpha_{3})}+c_{23}c_{21}^{*}e^{{\rm i}(\alpha_{3}-\alpha_{1})}+c_{31}c_{32}^{*}e^{{\rm i}(\alpha_{1}-\alpha_{2})}]
\end{eqnarray}
\end{widetext}

There are $C_{10}^{2}=45$ pair-wise superpositions in total in $|\psi_{3}^{3}\rangle_{\rm in}$. We can see from Eq.(\ref{eq3c2}) that only 18 pair-wise superpositions of them can make interference. One can check that each of the 18 superpositions has the local coherence of a single photon. The other 27 pair-wise superpositions in $|\psi_{3}^{3}\rangle_{\rm in}$ have only collective coherence and thus have no contribution to fringe visibility.
These superpositions include both the three-photon and two-photon superpositions between different paths,
such as $c_{1}|3_{\mathbf{k}_{1}}\rangle+c_{2}|3_{\mathbf{k}_{2}}\rangle$, $c_{1}|3_{\mathbf{k}_{1}}\rangle+c_{23}|2_{\mathbf{k}_{2}}\rangle|1_{\mathbf{k}_{3}}\rangle$,
and $c_{1}|3_{\mathbf{k}_{1}}\rangle+c_{21}|2_{\mathbf{k}_{2}}\rangle|1_{\mathbf{k}_{1}}\rangle=|1_{\mathbf{k}_{1}}\rangle(c_{1}|2_{\mathbf{k}_{1}}\rangle+c_{21}|2_{\mathbf{k}_{2}}\rangle)$,
$c_{12}|2_{\mathbf{k}_{1}}\rangle|1_{\mathbf{k}_{2}}\rangle+c_{23}|2_{\mathbf{k}_{2}}\rangle|1_{\mathbf{k}_{3}}\rangle=|1_{\mathbf{k}_{2}}\rangle(c_{12}|2_{\mathbf{k}_{1}}\rangle+c_{23}|1_{\mathbf{k}_{2}}\rangle|1_{\mathbf{k}_{3}}\rangle)$, etc.
In addition, following the method of the deduction in \ref{sec3b}, we have
\begin{widetext}
\begin{eqnarray}
\nonumber V_{3}^{3}&\leq& \frac{2\sqrt{3}}{3}\left\{|c_{1}c_{12}|+|c_{1}c_{13}|+|c_{2}c_{23}|+|c_{2}c_{21}|+|c_{3}c_{31}|+|c_{3}c_{32}|\right\}\\
\nonumber&+&\frac{2\sqrt{2}}{3}\left\{|c_{12}c_{33}|+|c_{21}c_{33}|+|c_{13}c_{33}|+|c_{31}c_{33}|+|c_{23}c_{33}|+|c_{32}c_{33}|\right\}\\
\nonumber &+&\frac{2}{3}\left\{2|c_{12}c_{21}|+2|c_{13}c_{31}|+2|c_{23}c_{32}|+|c_{12}c_{13}|+|c_{23}c_{21}|+|c_{31}c_{32}|\right\}.
\end{eqnarray}
\end{widetext}
Obviously, $V_{3}^{3}$ is generally less than the $l_1$-norm coherence of the input state $|\psi_{3}^{3}\rangle_{\rm in}$ in Eq.(\ref{eq35}). Through careful calculation
(including the consideration of the symmetry of the input states and the use of the extremum of binary functions given in Appendix B),
we find that when the superposition coefficients fulfill $c_{1}=c_{2}=c_{3}=\sqrt{3}/9$, $c_{12}=c_{23}=c_{31}=c_{21}=c_{32}=c_{13}=1/3$, $c_{33}=\sqrt{2}/3$, and the phase shifts fulfill $\alpha_{1}=4\pi/3$, $\alpha_{2}=2\pi/3$, $\alpha_{3}=0$,
the interference intensity of Eq.(\ref{eq3c1}) has the minimum $I_{\rm min}=0$ so that the visibility reaches the maximum $V_{3}^{3}=1$.

\section{Four-path interference}\label{sec4}
\subsection{Four-path 1-photon interference}
\label{sec4a}
For the four-path one-photon interference, we write the input state as
\begin{eqnarray}\label{eqf1}
  |\psi_{4}^{1}\rangle_{\rm in}=c_{1}|1_{\mathbf{k}_{1}}\rangle+c_{2}|1_{\mathbf{k}_{2}}\rangle+c_{3}|1_{\mathbf{k}_{3}}\rangle+c_{4}|1_{\mathbf{k}_{4}}\rangle
\end{eqnarray}
with the normalization $\sum_{j=1}^{4}|c_{j}|^{2}=1$. The output state after phase-shift operations reads
\begin{eqnarray}\label{eqf2}
\nonumber  |\psi_{4}^{1}\rangle_{\rm out}&=&c_{1}e^{{\rm i}\alpha_{1}}|1_{\mathbf{k}_{1}}\rangle+c_{2}e^{{\rm i}\alpha_{2}}|1_{\mathbf{k}_{2}}\rangle\\
  &+&c_{3}e^{{\rm i}\alpha_{3}}|1_{\mathbf{k}_{3}}\rangle+c_{4}e^{{\rm i}\alpha_{4}}|1_{\mathbf{k}_{4}}\rangle
\end{eqnarray}
Thus the intensity of the interference fringe is
\begin{eqnarray}\label{eqf3}
\nonumber I&=&|\varepsilon_{0}|^{2}\left\{1+2|c_{1}c_{2}|\cos\beta_{12}+2|c_{1}c_{3}|\cos\beta_{13}\right.\\
&+&2|c_{1}c_{4}|\cos\beta_{14}+2|c_{2}c_{3}|\cos(\beta_{13}-\beta_{12})\\
\nonumber&+&\left. 2|c_{2}c_{4}|\cos(\beta_{14}-\beta_{12})+2|c_{3}c_{4}|\cos(\beta_{14}-\beta_{13}) \right\},
\end{eqnarray}
where we denote $c_{j}=|c_{j}|e^{{\rm i}\delta_{j}}$ with $j=1,\ldots,4$, and $\beta_{ij}=(\alpha_{i}-\alpha_{j})+(\delta_{i}-\delta_{j})$. For the given input state $|\psi_{4}^{1}\rangle_{\rm in}$, we can always adjust the phase shifts
$\alpha_{j}$ with $j=1,\ldots,4$ to make $\beta_{12}=\beta_{13}=\beta_{14}=0$.
Thus the maximum of the interference intensity is given by
\begin{eqnarray}\label{eqf4}
\nonumber  I_{\rm max}&=&|\varepsilon_{0}|^{2}\left\{1+2|c_{1}c_{2}|+2|c_{1}c_{3}|+2|c_{1}c_{4}|\right.\\
  &+&\left.2|c_{2}c_{3}|+2|c_{2}c_{4}|+2|c_{3}c_{4}|\right\}.
\end{eqnarray}
The minimum of the interference intensity can be written as
\begin{eqnarray}\label{eqf5}
\nonumber I_{\rm min} &=&|\varepsilon_{0}|^{2}\left\{1+2|c_{1}c_{2}|\cos\beta_{12}^{0}+2|c_{1}c_{3}|\cos\beta_{13}^{0}\right.\\
&+&2|c_{1}c_{4}|\cos\beta_{14}^{0}+2|c_{2}c_{3}|\cos(\beta_{13}^{0}-\beta_{12}^{0})\\
\nonumber&+&\left. 2|c_{2}c_{4}|\cos(\beta_{14}^{0}-\beta_{12}^{0})+2|c_{3}c_{4}|\cos(\beta_{14}^{0}-\beta_{13}^{0}) \right\},
\end{eqnarray}
where $(\beta_{12}^{0},\beta_{13}^{0}, \beta_{14}^{0})$ denote the position for which the interference intensity reaches its minimum.
The fringe visibility $V_{4}^{1}$ in this case is then evaluated through Eq.(\ref{eq4}). We can make the following results:
1) The expression of $V_{4}^{1}$ includes all the pair-wise combinations of the coefficients in $|\psi_{4}^{1}\rangle_{\rm in}$: $|c_{1}c_{2}|$, $|c_{1}c_{3}|$, $|c_{1}c_{4}|$, $|c_{2}c_{3}|$, $|c_{2}c_{4}|$, $|c_{3}c_{4}|$,
meaning that the superposition of any two terms in Eq.(\ref{eqf1}) can make interference, or equivalently local coherence of a single photon can make interference. 2) One can easily find $V_{4}^{1}\leq 2|c_{1}c_{2}|+2|c_{1}c_{3}|+2|c_{1}c_{4}|+2|c_{2}c_{3}|+2|c_{2}c_{4}|+2|c_{3}c_{4}|\leq C(|\psi_{4}^{1}\rangle_{\rm in})$, i.e.,
visibility is generally less than the $l_1$-norm coherence of the corresponding input state. 3) In terms of the method of finding the extremum of multivariate function and after tedious calculations, we find that the maximal visibility of the interference fringe
in this case is also $V_{4}^{1}=1$, which is reached when $c_{1}=c_{2}=c_{3}=c_{4}=1/2$
and $\beta_{12}^{0}=0$, $\beta_{13}^{0}=\beta_{14}^{0}=\pi$ (or equivalently $\alpha_{1}=\alpha_{2}=0$ and
$\alpha_{3}=\alpha_{4}=-\pi$).

\subsection{Four-path 2-photon interference}
\label{sec4b}
The input state in this case can be written as
\begin{eqnarray}\label{eq33}
  |\psi_{4}^{2}\rangle_{\rm in}&=&c_{1}|2_{\mathbf{k}_{1}}\rangle+c_{2}|2_{\mathbf{k}_{2}}\rangle+c_{3}|2_{\mathbf{k}_{3}}\rangle+c_{4}|2_{\mathbf{k}_{4}}\rangle\\
  \nonumber &+&c_{12}|1_{\mathbf{k}_{1}}\rangle|1_{\mathbf{k}_{2}}\rangle+c_{13}|1_{\mathbf{k}_{1}}\rangle|1_{\mathbf{k}_{3}}\rangle+c_{14}|1_{\mathbf{k}_{1}}\rangle|1_{\mathbf{k}_{4}}\rangle\\
\nonumber  &+&c_{23}|1_{\mathbf{k}_{2}}\rangle|1_{\mathbf{k}_{3}}\rangle+c_{24}|1_{\mathbf{k}_{2}}\rangle|1_{\mathbf{k}_{4}}\rangle+c_{34}|1_{\mathbf{k}_{3}}\rangle|1_{\mathbf{k}_{4}}\rangle,
\end{eqnarray}
where the sum of the squares of the absolute values of all the complex coefficients equals one.
The intensity of the interference fringe can be written as
\begin{widetext}
\begin{eqnarray}\label{eq34}
\nonumber  I &=& 2|\varepsilon_{0}|^{2}\left\{1+ \sqrt{2}|c_{1}c_{12}|\cos\beta_{1}+\sqrt{2} |c_{1}c_{13}|\cos\beta_{2}+\sqrt{2}|c_{1}c_{14}|\cos\beta_{3}+|c_{12}c_{13}|\cos\beta_{4}+|c_{12}c_{14}|\cos\beta_{5}+|c_{13}c_{14}|\cos\beta_{6}\right.\\
 &+&\sqrt{2}|c_{2}c_{12}|\cos\beta_{7}+\sqrt{2} |c_{2}c_{23}|\cos\beta_{8}+\sqrt{2}|c_{2}c_{24}|\cos\beta_{9}+|c_{12}c_{23}|\cos\beta_{10}+|c_{12}c_{24}|\cos\beta_{11}+|c_{23}c_{24}|\cos\beta_{12}\\
\nonumber &+&\sqrt{2}|c_{3}c_{13}|\cos\beta_{13}+\sqrt{2} |c_{3}c_{23}|\cos\beta_{14}+\sqrt{2}|c_{3}c_{34}|\cos\beta_{15}+|c_{13}c_{23}|\cos\beta_{16}+|c_{13}c_{34}|\cos\beta_{17}+|c_{23}c_{34}|\cos\beta_{18}\\
\nonumber &+&\left. \sqrt{2}|c_{4}c_{14}|\cos\beta_{19}+\sqrt{2} |c_{4}c_{24}|\cos\beta_{20}+\sqrt{2}|c_{4}c_{34}|\cos\beta_{21}+|c_{14}c_{24}|\cos\beta_{22}+|c_{14}c_{34}|\cos\beta_{23}+|c_{24}c_{34}|\cos\beta_{24}\right\},
\end{eqnarray}
\end{widetext}
where $\beta_{j}$ with $j=1,\ldots,24$ are the functions of the phase shifts $\alpha_{i}$ with $i=1,\ldots,4$ and the arguments of the complex coefficients $c$'s in Eq.(\ref{eq33}).
Especially when all the coefficients $c$'s are positive (i.e. their arguments are zero), we have
\begin{eqnarray}
\nonumber\beta_{1}&=&\beta_{7}=\beta_{16}=\beta_{22}=\alpha_{1}-\alpha_{2},\\
\nonumber\beta_{2}&=&\beta_{10}=\beta_{13}=\beta_{23}=\alpha_{1}-\alpha_{3},\\
\nonumber\beta_{3}&=&\beta_{11}=\beta_{17}=\beta_{19}=\alpha_{1}-\alpha_{4},\\
\nonumber\beta_{4}&=&\beta_{8}=\beta_{14}=\beta_{24}=\alpha_{2}-\alpha_{3},\\
\nonumber\beta_{5}&=&\beta_{9}=\beta_{18}=\beta_{20}=\alpha_{2}-\alpha_{4},\\
\beta_{6}&=&\beta_{12}=\beta_{15}=\beta_{21}=\alpha_{3}-\alpha_{4}.
\end{eqnarray}
Obviously, $\beta_{j}$ are generally not completely independent. For the given input state of Eq.(\ref{eq33}), we can adjust the phase shifts  $\alpha_{i}$ with $i=1,\ldots,4$ to make the interference intensity
Eq.(\ref{eq34}) reaching its maximum $I_{\rm max}$ or minimum $I_{\rm min}$. Then the visibility $V_{4}^{2}$ can be evaluated via Eq.(\ref{eq4}). Though the calculation is very complicated, we can now find the following results:

1) There are $C_{10}^{2}=45$ pair-wise superpositions in $|\psi_{4}^{2}\rangle_{\rm in}$, in which 24 superpositions can make interference and the rest 21 superpositions have no contribution to fringe visibility.
One can check that the superpositions that can make interference must contain the local coherence of a single photon. Conversely, the superpositions that have no contribution to visibility have only collective coherence,
such as $c_{1}|2_{\mathbf{k}_{1}}\rangle+c_{2}|2_{\mathbf{k}_{2}}\rangle$, $c_{1}|2_{\mathbf{k}_{1}}\rangle+c_{23}|1_{\mathbf{k}_{2}}\rangle|1_{\mathbf{k}_{3}}\rangle$
and $c_{12}|1_{\mathbf{k}_{1}}\rangle|1_{\mathbf{k}_{2}}\rangle+c_{34}|1_{\mathbf{k}_{3}}\rangle|1_{\mathbf{k}_{4}}\rangle$, etc.

2)Following the method of the deduction in \ref{sec3b}, we find that the visibility of interference fringe fulfills
\begin{widetext}
\begin{eqnarray}\label{eq34a}
\nonumber  V_{4}^{2} &\leq & \sqrt{2}\left\{|c_{1}c_{12}|+|c_{1}c_{13}|+|c_{1}c_{14}|+|c_{2}c_{12}|+|c_{2}c_{23}|+|c_{2}c_{24}|\right\}\\
 &+&\sqrt{2}\left\{|c_{3}c_{13}|+|c_{3}c_{23}|+|c_{3}c_{34}|+|c_{4}c_{14}|+|c_{4}c_{24}|+|c_{4}c_{34}|\right\}\\
\nonumber &+&|c_{12}c_{13}|+|c_{12}c_{14}|+|c_{13}c_{14}|+|c_{12}c_{23}|+|c_{12}c_{24}|+|c_{23}c_{24}|\\
\nonumber &+&|c_{13}c_{23}|+|c_{13}c_{34}|+|c_{23}c_{34}|+|c_{14}c_{24}|+|c_{14}c_{34}|+|c_{24}c_{34}|.
\end{eqnarray}
\end{widetext}
Obviously, $V_{4}^{2}$ is generally less than the $l_1$-norm coherence of the input state $|\psi_{4}^{2}\rangle_{\rm in}$.

3) Through careful and tedious calculations, we find when all the superposition coefficients are positive and fulfill $c_{1}=c_{2}=c_{3}=c_{4}=1/4$, $c_{12}=c_{13}=c_{14}=c_{23}=c_{24}=c_{34}=1/\sqrt{8}$, and the phase shifts fulfill
$\alpha_{1}=\alpha_{2}=0$, $\alpha_{3}=\alpha_{4}=-\pi$, then the interference intensity of Eq.(\ref{eq34}) equals zero, meaning that the fringe visibility can reach the maximum $V_{4}^{2}=1$.

\subsection{Four-path 3-photon interference}
\label{sec4c}
The input state is
\begin{widetext}
\begin{eqnarray}
\label{eq4c1}
\nonumber  |\psi_{4}^{3}\rangle_{\rm in}&=&c_{1}|3_{\mathbf{k}_{1}}\rangle+c_{2}|3_{\mathbf{k}_{2}}\rangle+c_{3}|3_{\mathbf{k}_{3}}\rangle+c_{4}|3_{\mathbf{k}_{4}}\rangle\\
 &+&c_{12}|2_{\mathbf{k}_{1}}\rangle|1_{\mathbf{k}_{2}}\rangle+c_{13}|2_{\mathbf{k}_{1}}\rangle|1_{\mathbf{k}_{3}}\rangle+c_{14}|2_{\mathbf{k}_{1}}\rangle|1_{\mathbf{k}_{4}}\rangle
  +c_{21}|2_{\mathbf{k}_{2}}\rangle|1_{\mathbf{k}_{1}}\rangle+c_{23}|2_{\mathbf{k}_{2}}\rangle|1_{\mathbf{k}_{3}}\rangle+c_{24}|2_{\mathbf{k}_{2}}\rangle|1_{\mathbf{k}_{4}}\rangle\\
\nonumber  &+&c_{31}|2_{\mathbf{k}_{3}}\rangle|1_{\mathbf{k}_{1}}\rangle+c_{32}|2_{\mathbf{k}_{3}}\rangle|1_{\mathbf{k}_{2}}\rangle+c_{34}|2_{\mathbf{k}_{3}}\rangle|1_{\mathbf{k}_{4}}\rangle
  +c_{41}|2_{\mathbf{k}_{4}}\rangle|1_{\mathbf{k}_{1}}\rangle+c_{42}|2_{\mathbf{k}_{4}}\rangle|1_{\mathbf{k}_{2}}\rangle+c_{43}|2_{\mathbf{k}_{4}}\rangle|1_{\mathbf{k}_{3}}\rangle\\
\nonumber  &+&c_{123}|1_{\mathbf{k}_{1}}\rangle|1_{\mathbf{k}_{2}}\rangle|1_{\mathbf{k}_{3}}\rangle
 +c_{124}|1_{\mathbf{k}_{1}}\rangle|1_{\mathbf{k}_{2}}\rangle|1_{\mathbf{k}_{4}}\rangle+c_{134}|1_{\mathbf{k}_{1}}\rangle|1_{\mathbf{k}_{3}}\rangle|1_{\mathbf{k}_{4}}\rangle
 +c_{234}|1_{\mathbf{k}_{2}}\rangle|1_{\mathbf{k}_{3}}\rangle|1_{\mathbf{k}_{4}}\rangle,
\end{eqnarray}
\end{widetext}
where the sum of the squares of the absolute values of all the complex coefficients equals one. As the calculation in this case is very complicated, we only give the main results as follows.

1) There are $C_{20}^{2}=190$ pair-wise superpositions in $|\psi_{4}^{3}\rangle_{\rm in}$, in which 60 superpositions can make interference and the rest 130 superpositions have no contribution to fringe visibility.
The pair-wise superpositions that can make interference correspond to respectively the pair-wise coefficient combinations in Eq(\ref{eq4c2})(see below). One can check that these superpositions contain the local coherence of a single photon. Conversely, the superpositions that have no contribution to visibility have only collective coherence.

2) Following the method of the deduction in \ref{sec3b}, we have
\begin{widetext}
\begin{eqnarray}\label{eq4c2}
\nonumber  V_{4}^{3} &\leq & \frac{2}{3}\sqrt{3}\left\{|c_{1}c_{12}|+|c_{1}c_{13}|+|c_{1}c_{14}|+|c_{2}c_{21}|+|c_{2}c_{23}|+|c_{2}c_{24}|\right\}\\
\nonumber &+&\frac{2}{3}\sqrt{3}\left\{|c_{3}c_{31}|+|c_{3}c_{32}|+|c_{3}c_{34}|+|c_{4}c_{41}|+|c_{4}c_{42}|+|c_{4}c_{43}|\right\}\\
\nonumber &+&\frac{2}{3}\sqrt{2}\left\{|c_{12}c_{123}|+|c_{12}c_{124}|+|c_{21}c_{123}|+|c_{21}c_{124}|+|c_{13}c_{123}|+|c_{13}c_{134}|+|c_{31}c_{123}|+|c_{31}c_{134}|\right\}\\
 &+&\frac{2}{3}\sqrt{2}\left\{|c_{14}c_{124}|+|c_{14}c_{134}|+|c_{41}c_{124}|+|c_{41}c_{134}|+|c_{23}c_{123}|+|c_{23}c_{234}|+|c_{32}c_{123}|+|c_{32}c_{234}|\right\}\\
\nonumber &+&\frac{2}{3}\sqrt{2}\left\{|c_{24}c_{124}|+|c_{24}c_{234}|+|c_{42}c_{124}|+|c_{42}c_{234}|+|c_{34}c_{134}|+|c_{34}c_{234}|+|c_{43}c_{134}|+|c_{43}c_{234}|\right\}\\
\nonumber &+&\frac{4}{3}\left\{|c_{12}c_{21}|+|c_{13}c_{31}|+|c_{14}c_{41}|+|c_{23}c_{32}|+|c_{24}c_{42}|+|c_{34}c_{43}|\right\}\\
\nonumber&+&\frac{2}{3}\left\{|c_{12}c_{13}|+|c_{12}c_{14}|+|c_{13}c_{14}|+|c_{21}c_{23}|+|c_{21}c_{24}|+|c_{23}c_{24}|+|c_{31}c_{32}|+|c_{31}c_{34}|+|c_{32}c_{34}|\right\}\\
\nonumber&+&\frac{2}{3}\left\{|c_{41}c_{42}|+|c_{41}c_{43}|+|c_{42}c_{43}|+|c_{123}c_{124}|+|c_{123}c_{134}|+|c_{124}c_{134}|+|c_{123}c_{234}|+|c_{124}c_{234}|+|c_{134}c_{234}|\right\}.
\end{eqnarray}
\end{widetext}
Obviously, $V_{4}^{3}$ is generally less than the $l_1$-norm coherence of the input state $|\psi_{4}^{3}\rangle_{\rm in}$.

3) Through careful and tedious calculations, we find when all the superposition coefficients are positive and fulfill $c_{1}=c_{2}=c_{3}=c_{4}=1/8$, $c_{12}=c_{13}=c_{14}=c_{21}=c_{23}=c_{24}=c_{31}=c_{32}=c_{34}=c_{41}=c_{42}=c_{43}=\sqrt{3}/8$,
$c_{123}=c_{124}=c_{134}=c_{234}=\sqrt{6}/8$, and the phase shifts fulfill
$\alpha_{1}=\alpha_{2}=0$, $\alpha_{3}=\alpha_{4}=-\pi$, then the fringe visibility reaches the maximum $V_{4}^{3}=1$.

\section{L-path interference}
\label{sec5}
For the L-path interference, we only discuss the case of a single input photon. The input state is given by
\begin{eqnarray}
  |\psi_{L}^{1}\rangle_{\rm in}=c_{1}|1_{\mathbf{k}_{1}}\rangle+c_{2}|1_{\mathbf{k}_{2}}\rangle+\cdots+c_{L}|1_{\mathbf{k}_{L}}\rangle
\end{eqnarray}
with the normalization $\sum_{j=1}^{L}|c_{j}|^{2}=1$. The output state reads
\begin{eqnarray}
\nonumber  |\psi_{L}^{1}\rangle_{\rm out}=c_{1}e^{{\rm i}\alpha_{1}}|1_{\mathbf{k}_{1}}\rangle+c_{2}e^{{\rm i}\alpha_{2}}|1_{\mathbf{k}_{2}}\rangle+\cdots+c_{L}e^{{\rm i}\alpha_{L}}|1_{\mathbf{k}_{L}}\rangle
\end{eqnarray}
The intensity of the interference fringe is
\begin{equation}
  I=|\varepsilon_{0}|^{2}\left\{1+2\sum_{i<j}|c_{i}c_{j}|\cos\beta_{ij}\right\},
\end{equation}
where we denote $c_{j}=|c_{j}|e^{{\rm i}\delta_{j}}$ with $j=1,\ldots,L$, and $\beta_{ij}=(\alpha_{i}-\alpha_{j})+(\delta_{i}-\delta_{j})$, the summation is for both $i$ and $j$ with $i<j$.
For the given input state $|\psi_{L}^{1}\rangle_{\rm in}$, we can always adjust the phase shifts
$\alpha_{j}$ with $j=1,\ldots,L$ to make all of the $\beta_{ij}$'s to be zero.
Thus the maximum of the interference intensity is given by
\begin{equation}
  I_{\rm max}=|\varepsilon_{0}|^{2}[1+2\sum_{i<j}|c_{i}c_{j}|].
\end{equation}
The minimum of the interference intensity may be written as
\begin{equation}
  I_{\rm min}=|\varepsilon_{0}|^{2}\left[1+2\sum_{i<j}|c_{i}c_{j}|\cos\beta_{ij}^{0}\right],
\end{equation}
where $\beta_{ij}^{0}$'s denote the position of $I_{\rm min}$.
The visibility of the interference fringe is thus given by
\begin{equation}
  V_{L}^{1}=\frac{\sum_{i<j}|c_{i}c_{j}|(1-\cos\beta_{ij}^{0})}{1+\sum_{i<j}|c_{i}c_{j}|(1+\cos\beta_{ij}^{0})}.
\end{equation}
From this result, we find that $V_{L}^{1}\leq 2\sum_{i<j}|c_{i}c_{j}|$, i.e., $V_{L}^{1}$ is generally less the $l_1$-norm coherence of the input state $|\psi_{L}^{1}\rangle_{\rm in}$.
Unfortunately, we can not demonstrate in this universal case (i.e.,arbitrary finite $L$) whether the maximal visibility of the interference fringe can reach one.

\section{conclusions}
\label{sec6}
In conclusion, based on the concept of pair-wise coherence, we have studied the relation between the $l_1$ norm of coherence and the quantum interference in Young's multi-path experiments for the cases that the input photons can be entangled each other. Three main results have been found. Firstly, the local coherence of each single photon can make quantum interference, but the collective coherence between photons has no contribution to quantum interference. Secondly, the visibility of interference fringe is generally less than the $l_1$ norm of coherence of the corresponding input states, which suggests that the $l_1$ norm of coherence is only the necessary but not sufficient condition of quantum interference. Finally, we have found that the maximal visibility of the interference fringe can reach one, and the corresponding optimal input states for producing this maximal visibility of interference fringe have been presented for the considered several cases. For the one-photon interference (including both two paths and multiple paths), the equal weight superposition among all the paths is the optimal input state. For the multi-photon interference, the optimal input state depends on the number of both photons and paths, and no universal rule has been found.

It is worthwhile to point out that the special input states for realizing the desired multi-photon interference can not be generated in general by simply pushing photons through pinholes. Some particular devices must be engineered for generating these states. For example, the methods for producing the well-known N00N states for photons\cite{Fiurasek2002,Kok2002}, as well as for atoms \cite{Zhong2010} and trapped ions \cite{Huang2020}, have been proposed, deliberately aiming at the very important applications in quantum lithography\cite{Boto2000}, quantum metrology\cite{Bollinger1996}, quantum cryptography \cite{Tittel2000} and quantum teleportation \cite{Riedmatten2004}.

Quantum coherence and quantum interference are related closely. Recently, how $l_1$ norm of coherence is connected to quantum interference has been studied \cite{Biswas2017,Qureshi2017,Paul2017}. Especially, the authors in Reference \cite{Wang2017} have demonstrated experimentally the consistency between $l_1$ norm of coherence and quantum interference fringes. However, our research suggests that, for the multi-photon Young's experiments, it is the local (single photon) coherence, not the collective coherence (i.e, entanglement between photons) that is a good manifestation of the quantum interference. It means that in the level of single photon, the interference fringes is a good manifestation of the $l_1$ norm of coherence. But for multi-photon system, it is may not the case: There are some multi-photon states (such as N00N states), which have $l_1$ norm of coherence but does not take place quantum interference.

\begin{acknowledgements}
{This work is supported by the National Natural Science Foundation of China (Grant Nos.1217050862, 11275064).}
\end{acknowledgements}

\renewcommand{\theequation}{A.\arabic{equation}}
\setcounter{equation}{0}
\section*{Appendix A: Lagrange multiplier}
In this Appendix, we present the outline for using the Lagrange multiplier to find the conditional extremum of a given function.
To find the extremum of the function $f(x_{1}, x_{2}, \ldots, x_{n})$ with $n$ variables under the restricted conditions
\begin{equation}\label{app1}
  g_{i}(x_{1}, x_{2}, \ldots, x_{n})=0
\end{equation}
with $i=1, 2,\ldots, k$ ($k<n$), one may construct the Lagrange function
\begin{eqnarray}\label{app2}
\mathcal{L}(x_{1},\ldots, x_{n}; \lambda_{1}, \ldots, \lambda_{k})&=&f(x_{1}, \ldots, x_{n})\\
\nonumber  &-&\sum_{i=1}^{k}\lambda_{i}g_{i}(x_{1}, \ldots, x_{n}),
\end{eqnarray}
with $\lambda_{i}$ the Lagrange  multipliers. The following equations
\begin{equation}\label{app3}
 \begin{array}{cc}
   \frac{\partial \mathcal{L}}{\partial x_{j}}=0, &  (j=1,\ldots, n)
 \end{array}
  \end{equation}
plus the restricted conditions of Eq.(\ref{app1}) together forms a complete set of equations with respect to $x_{1},\ldots, x_{n}, \lambda_{1}, \ldots, \lambda_{k}$, which solutions give the possible positions of the extremum of function $f(x_{1}, x_{2}, \ldots, x_{n})$.

\renewcommand{\theequation}{B.\arabic{equation}}
\setcounter{equation}{0}
\section*{Appendix B: Extremum of binary functions}
In this Appendix, we present the outline for calculating the extremum of a given binary function. Given a binary function $f(x,y)$, the possible extreme point $(x_{0},y_{0})$ fulfills
\begin{equation}\label{app4}
  \begin{array}{cc}
    f'_{x}(x_{0},y_{0})=0, & f'_{y}(x_{0},y_{0})=0.
  \end{array}
\end{equation}
Denote $A=f''_{xx}(x_{0},y_{0})$, $B=f''_{xy}(x_{0},y_{0})$ and $C=f''_{yy}(x_{0},y_{0})$. If $B^{2}-AC<0$ and $A>0$, then $ f(x_{0},y_{0})$ is the minimum; If $B^{2}-AC<0$ and $A<0$, then $ f(x_{0},y_{0})$ is the maximum.
If $B^{2}-AC>0$, then $ f(x_{0},y_{0})$ is not the extreme. If $B^{2}-AC=0$, we can not determine whether  $ f(x_{0},y_{0})$ is extreme.

\end{document}